\documentclass[12pt,preprint]{aastex}
 
\shorttitle{Reionization from GRBs}
\shortauthors{Murakami et al.}

\begin{document}


\title{
The Reionization History and Early Metal Enrichment 
inferred from the Gamma-Ray Burst Rate
}

\author{T. Murakami\altaffilmark{1} and D. Yonetoku}
\affil{Faculty of Science, University of Kanazawa, Kanazawa, Ishikawa
       920-1192 Japan}
\email{murakami@astro.s.kanazawa-u.ac.jp, yonetoku@astro.s.kanazawa-u.ac.jp}
\author{M. Umemura }
\affil{Center for Computational Sciences, University of Tsukuba,
       Ibaraki, 305-8577 Japan}
\email{umemura@rccp.tsukuba.ac.jp}
\author{T. Matsubayashi}
\affil{Ebisuzaki Computational Astrophysics, Riken, Wako, Saitama
      152-8551 Japan}
\email{tatsushi@riken.jp}
\and
\author{R. Yamazaki\altaffilmark{2}}
\affil{Department of Earth and Space Science, Osaka-University, Osaka,
          560-0043 Japan}
\email{ryo@vega.ess.sci.osaka-u.ac.jp}
\altaffiltext{1}{Visiting Professor, Institute of Space and
 Astronautical Science, Sagamihara, Kanagawa 229-0012 Japan}
\altaffiltext{2}{PHD Fellow, JSPS}

\begin{abstract}

Based on the gamma-ray burst (GRB) event rate at redshifts of $4 \leq
z \leq 12$, which is assessed by the spectral peak
energy-to-luminosity relation recently found by Yonetoku et al., we
observationally derive the star formation rate (SFR) for Pop III stars
in a high redshift universe. As a result, we find that Pop III stars
could form continuously at $4 \leq z \leq 12$.  Using the derived Pop
III SFR, we attempt to estimate the ultraviolet (UV) photon emission
rate at $7 \leq z \leq 12$ in which redshift range no observational
information has been hitherto obtained on ionizing radiation
intensity.  We find that the UV emissivity at $7 \leq z \leq 12$ can
make a noticeable contribution to the early reionization.  The maximal
emissivity is higher than the level required to keep ionizing the
intergalactic matter at $7 \leq z \leq 12$.  However, if the escape
fraction of ionizing photons from Pop III objects is smaller than
10\%, then the IGM can be neutralized at some redshift, which may lead
to the double reionization.  As for the enrichment, the ejection of
all metals synthesized in Pop III objects is marginally consistent
with the IGM metallicity, although the confinement of metals in Pop
III objects can reduce the enrichment significantly.

\end{abstract}


\section{Introduction}

The recent observation on the cosmic microwave background by {\it
Wilkinson Microwave Anisotropy Probe} ({\it WMAP}) has revealed that
the dark age in the universe ended quite early by the reionization at
$z_r \sim 20^{+10}_{-9}$ \citep{ben03,spe03}.  But, the origin of
ionizing radiation in such an early epoch and the subsequent
ionization history after the early reionization are not elucidated
very well.  The first possibility for the ionizing source is Pop III
stars \citep{Cen03a,Cia03, Wyi03, Somer03, FK03}, and the other
possibilities are mini quasars \citep{Mad04} or BH accretion
\citep{SU96, RO04b}.  The early reionization by Pop III stars is
recently under debate.  \citet{sok04} argue that the reionization by
Pop III stars is marginally compatible with the Thomson optical depth
measured by the {\it WMAP}, $\tau_e =0.17 \pm 0.04$, even if the
escape fraction of ionizing photons from early formed objects is
unity. In addition, \citet{RO04a} explore the feedback effect by
supernova (SN) energy input in Pop III objects, and find that the
suppression of Pop III star formation by SNe is so intensive that Pop
III stars cannot contribute significantly to the cosmic reionization.
The strong suppression of Pop III star formation or the early shift to
Pop II or Pop I star formation may bring the vanishment of ionizing
radiation, resulting in the re-neutralization of intergalactic medium
(IGM).  This may lead to the so-called double reionization
\citep{Cen03b, Wyi04}.

Also, Pop III SNe can be the origin of IGM heavy elements
\citep{nak01,scan02,RO04a}.  The detailed analysis on quasar metal
absorption lines have shown that the IGM is enriched by heavy elements
from a level of $\sim$ 10$^{-3} Z_\odot$ at $z \la 4.3$ to $\sim$
10$^{-4} Z_\odot$ at $z \sim 5.3$, where $Z_\odot$ is the Solar
abundance \citep{cow98,son01}.  It is quite perplexing that heavy
elements are observed commonly and uniformly in the low density space
where no bright galaxies are found.  Although the outflow from small
proto-galaxies could be responsible for the enrichment of IGM
\citep{Mad01,Mori02}, the much earlier enrichment by Pop III objects
may provide a potential solution for the uniformly distributed
metagalactic heavy elements.  Hence, Pop III star formation history in
high redshifts could be a key for the early enrichment.

We cannot observe directly Pop III stars at $7<z< z_r$, but GRBs at
high redshifts can be a potential tool to probe the formation history
of Pop III stars.  Since the discovery of X-ray afterglow from GRB
970228 \citep{cos97,van97}, the distances to bright GRBs are known to
be cosmological \citep{met97,gre04}.  The most distant GRB hitherto
identified spectroscopically is GRB~000131 at $z=4.5$
\citep{and00,gre04}.  However, there are numerous much weaker GRBs
without known distance, listed in the archive by the {\it Burst And
Transient Source Experiment (BATSE)} on-board the Compton gamma-ray
astronomy satellite \citep{bat04}.  Recently, \citet{yon04} estimated
distances of 689 BATSE weak GRBs, based on the spectral peak
energy-to-luminosity relation \citep{ama02, yon04}, and found that
most of GRBs are more remote up to $z \sim 12$.  A recent great
advance on GRBs is the finding of the obvious association between
GRB~030329 and SN~2003dh \citep{hjo03,gre03,mat03,pri03,ume03}. This
association firmly established the connection between a GRB and an
energetic supernova (hypernova), which is the collapse of a massive
star \citep{woo99,pac98}.  Since Pop III stars are likely to form in a
top-heavy initial mass function (IMF) \citep{BCL99, Abel00, Abel02,
nak01}, a portion of GRBs at high redshifts possibly result from Pop
III hypernovae. Actually, \citet{heg03} estimated the fraction of Pop
III GRBs using the IMF by Nakamura \& Umemura, and found that several
percent of Pop III stars result in GRBs.

In this Letter, we derive the star formation rate (SFR) of Pop III
stars observationally for the first time out to $z \sim 12$ by use of
the absolute GRB event rate.  Then, using the Pop III SFR, we assess
the UV photon emission rate at $7 \leq z \leq 12$, in which redshift
range no observational constraint has been hitherto placed on ionizing
radiation intensity.  Also, we analyze the early metal enrichment of
IGM by the derived Pop III SFR.  Here, we assume a $\Lambda$CDM
universe with $\Omega_\Lambda = 0.7$, $\Omega_m = 0.3$ and $H_{0}=
72~{\rm km s^{-1} Mpc^{-1}}$ \citep{ben03,spe03}.

\section{GRB Rate and Pop III SFR}

\subsection{ Absolute GRB Event Rate }

\citet{yon04} published the relative value of the GRB event rate up to
$z \sim 12$, based on the spectral peak energy-to-peak luminosity
relation.  In this Letter, we need to estimate an absolute value of
GRB event rate ($\dot{\rho}_{\rm GRB}$) in the early universe. In
order to do that, it is necessary to make corrections for
jet-collimation, because the GRB emission is collimated in the order
of 0.1 radian \citep{kul99,fra01}.  If the reported evolution in
luminosity \citep{yon04} can be attributed to the evolution of
jet-collimation, we can correct the collimation using the observed
evolution in luminosity with the index $\kappa$ defined as
(1+z)${^{2.6\kappa}}$, where 2.6 represents the observed evolution in
luminosity \citep{yon04}.  After the correction against
jet-collimation, the absolute event rate $\dot{\rho}_{\rm GRB}$ in a
comoving volume is obtained.  It is not quite certain whether the
observed evolution in luminosity is fully due to the collimation or
partially due to intrinsic evolution in luminosity.  So, we use the
intermediate case of $\kappa$=0.5 for simplicity in this paper.  The
resultant rate in units of Gpc${^{-3}}$ yr${^{-1}}$ 
is shown in Fig. 1 together with the observed data with
the systematic and statistical errors.  The typical rate at $z \sim
12$ is roughly 10${^{4}}$ Gpc${^{-3}}$ yr${^{-1}}$ for the case of
$\kappa$=0.5.

\subsection{Pop III SFR}

The absolute GRB rate allows us to estimate the SFR of Pop III stars
with the assistance of the Pop III IMF.  Because of the lack of heavy
elements, the IMF for Pop III stars is likely to be top heavy with massive
stars of $\ga 50M_{\odot}$, in contrast to the present-day stars
\citep{Abel02,nak01,TM04}.  The possible formation of even lower mass
Pop III stars with 1 to 50 M$_{\odot}$ is also pointed out
\citep{nak01}.  But these stars might be a quite small contribution to
the UV emission rate or GRB event rate, and therefore omitted in the
present analysis.

The theory on the stellar evolution of extremely metal poor ($\la
10^{-4}Z_\odot$) stars \citep{heg03} shows that the stars with about 8
to 40 M$_{\odot}$ will evolve and explode as type-II SNe, leaving a
neutron star or a black hole and supplying heavy elements into space.
The stars with 40 to 140 $M_{\odot}$ will produce UV photons during
their evolutionary pass, but will not supply any heavy element into
space, because they will collapse completely, forming a black hole.
The stars with 140 to 260 M$_{\odot}$ will explode as a
pair-instability SNe, forming a black-hole and supplying heavy
elements into space.  Among these stars, the key interest is the stars
with 100 to 140 $M_{\odot}$, which can result in GRBs as progenitors.
This mass range is sometimes called an energetic SN
(hypernova). Hence, in metal poor environments, GRB events occur only
in the narrow mass range in the Pop III IMF.

Here, we employ the Pop III IMF by \citet{nak01}, which is
in the form of a double-peaked function with sharp cutoffs 
with power-law shape:
$\alpha=\beta=1.35$ and cutoffs: $m_{\rm p1}=1M_\odot$, and
$m_{\rm p2}=50M_\odot$ respectively (see Nakamura \& Umemura 2001 for
details).  As stated above, we consider only a high-mass part of the
IMF.  The normalization factor can be determined by the GRB event
rate, $\dot{\rho}_{\rm GRB}$, which corresponds to a mass range of
$100-140 M_{\odot}$.  Then, if the Pop III fraction $f_{\rm GRB}^{\rm
III}$ in the observed GRB rate is given, we can derive the Pop III SFR
in terms of
\begin{equation}
\dot{\rho}_{\rm *}^{\rm III} \simeq 3\times10^{2}~f_{\rm GRB}^{\rm III}
\dot{\rho}_{\rm GRB} M_{\odot} {\rm Gpc^{-3} yr^{-1}}
\label{SFR}
\end{equation}
after an integration over the IMF.  Actually, $f_{\rm GRB}^{\rm III}$
is a function of time.  \citet{heg03} show that the GRB range quickly
expands for $Z>10^{-4}Z_\odot$ (Pop I/II), but simultaneously the IMF
is likely to become the present-day type (Salpeter) IMF for the
enriched gas \citep{bromm01}.  As a result, the mass fraction of GRBs
originating from Pop I/II stars is roughly twice as high as that of
Pop III GRBs. Hence, if $f_*^{\rm III}$ is the fraction of Pop III
stars, the $f_{\rm GRB}^{\rm III}$ is given by
\begin{equation}
f_{\rm GRB}^{\rm III}
\simeq 0.5 f_*^{\rm III} /(1-0. 5 f_*^{\rm III}).
\label{fGRB}
\end{equation}
It is expected that $f_*^{\rm III}$ is near unity in an early epoch,
and then decreases with the metal enrichment.  \citet{scan03} have
derived the $f_*^{\rm III}$ as a function of redshift.
Theyhave analyzed the dependence of $f_*^{\rm III}$ upon the total energy
input into outflows from Pop III ejecta per unit gas mass, $E_g^{\rm
III}$ in units of $10^{51}$~ergs $M_\odot$$^{-1}$. 
The model with $E_g^{\rm III}=10^{-4.5}$ is roughly corresponding
to $f_*^{\rm III}=40-50\%$, while the model with 
$E_g^{\rm III}=10^{-3.5}$ is to $f_*^{\rm III}=10\%$.
Hence, combining (\ref{fGRB}) with (\ref{SFR}), it is concluded 
that the observed GRB event rate implies the continuous formation
of Pop III stars at $4 \leq z \leq 12$.

\section{Reionization and Enrichment}

\subsection{UV Emissivity and Reionization History}

The massive stars with more than 50 $M_{\odot}$ radiate photons in the
black body of $\sim 10^5$ K, which are energetic enough to ionize
neutral hydrogen. Their luminosities should be near the Eddington
luminosity \citep{sch02}, so $\sim 10^{62} (M/M_{\odot})$ UV photons
are emitted during their lifetime of $\sim 10^{6}$ yrs.  For the Pop
III SFR assessed above, we can evaluate the total UV photon emission
rate per unit cosmological comoving volume as
\begin{equation}
\dot{\cal{N}}_\gamma \simeq 
3 \times 10^{62} \dot{\rho}_{\rm *}^{\rm III} ~{\rm Gpc^{-3} yr^{-1}}
\end{equation}
for stars heavier than 50 $M_{\odot}$, where
$\dot{\rho}_{\rm *}^{\rm III}$ is given in units of 
$M_{\odot} {\rm Gpc^{-3} yr^{-1}}$.
The resultant UV emissivity by Pop III stars is shown in Fig. 2.  
On the other hand, the UV photon emission rate required 
to keep ionizing hydrogen in the universe is estimated as
$Cn_{H}(z=0)/t_{rec}$ \citep{mad99}, where $C$ is 
the clumping factor of IGM
and $t_{rec}$ is the recombination time, $t_{rec} = 1100 
(1+z)^{-3}(\Omega_b h^{2}/0.0224)^{-1}~{\rm Gyr}$.  
Several authors have argued the effect of clumpiness 
on the reionization \citep{GO97, NUS01, SH03}.
Since the clumpiness shortens the recombination time
in HII regions, then the clumping factor should be estimated
by $C = \langle n_{\rm HII}^2 \rangle / \langle n_{\rm HII} \rangle^2 $.
When all the matter is totally ionized, $C$ becomes equal to 
the clumping factor of baryons, 
$C = \langle n_{b}^2 \rangle / \langle n_{b} \rangle^2 $.
But, if the IGM is partially ionized 
where density peaks are self-shielded, $C$ is smaller than 
$\langle n_{b}^2 \rangle / \langle n_{b} \rangle^2 $. 
Therefore, $C=1$ corresponds to
the minimum level requisite for the IGM ionization, while
$C = \langle n_{b}^2 \rangle / \langle n_{b} \rangle^2 $
does to the maximum level. 
In Fig. 2, the UV emission rate for reionization is shown for $C=1$ 
and $C = \langle n_{b}^2 \rangle / \langle n_{b} \rangle^2 $ which
is taken from \citet{GO97}.
The most recent results of WMAP suggest much higher clumping factor
than their value but we should wait for the second WMAP results for
more accurate disccusion.  As seen in Fig. 2, the UV photon emission
rate inferred from the GRB events exceeds the maximum level for the
IGM ionization at all epochs of $7 \leq z \leq 12$, for the $E_g^{\rm
III}=10^{-4.5}$ model or for the Pop III GRB fraction larger than
several 10\%.  However, if the escape fraction of ionizing photons is
much smaller than 10\%, the photon emission rate can fall below the maximum
rate.  Especially, when larger systems form at lower redshifts in a
CDM universe, the escape fraction may be reduced significantly
\citep{Kitayama04,Whalen04}.  Then, the IGM may be neutralized at some
redshift, At lower redshifts, Pop I/II stars would make more
contributions to ionizing photon budget, which may reionize the IGM
again.

\subsection{Early Enrichment of IGM}

Next, we consider the enrichment by heavy elements synthesized in Pop
III objects.  Since the amount of mass ejection from energetic SNe
(hypernovae) and from pair instability SNe into space is very roughly
$\sim$ 0.1 and $\sim$ 0.5 of the initial mass for $100-140 M_{\odot}$
and $140-260 M_{\odot}$ respectively \citep{heg03}, we can estimate
the heavy element production rate into the IGM as $0.3 \dot{\rho}_{\rm
*}^{\rm III}$ $M_{\odot}$ Gpc${^{-3}}$ yr${^{-1}}$. Using the Pop III
SFR, the metallicity of the IGM enriched by Pop III stars is estimated
to be $2.2 \times 10^{-4} Z_\odot$ for $E_g^{\rm III}=10^{-3.5}$ and
$1.2 \times 10^{-3} Z_\odot$ for $E_g^{\rm III}=10^{-4.5}$.  
The model with $E_g^{\rm III}=10^{-4.5}$ appears to overproduce the IGM metals,
if compared to $\sim$ 10$^{-4} Z_\odot$ at $z \sim 5.3$ \citep{son01}.
However, it should be noted that all synthesized metals 
cannot be spread out from Pop III objects \citep{Norman04}. 
Then, even the model with $E_g^{\rm III}=10^{-4.5}$ may not
provide sufficient metals for the IGM metallicity.

\section{Conclusions and Discussion}

In this paper, we have estimated the absolute GRB event rate out to $z
\sim $12 observationally with the correction for collimation.  There
have been also several efforts to estimate a GRB event rate in the
early universe, using the independent method
\citep{mur03,fen00,sch01}. All the results are broadly consistent with
each other and show that there is no break in the GRB event rate as a
function of redshift.

Based on the GRB event rate, we have found that Pop III stars could
form continuously at $4 \leq z \leq 12$. This implies that Pop III
star formation is not likely to be strongly suppressed in early
epochs.  This is qualitatively consistent with the recent study on the
star formation history in dwarf spheroidals in the Local group, which
shows the continuous star formation even after the reionization
\citep{Grebel04}.  Also in recent theoretical studies, it is shown
that dwarf galaxies can form owing to the self-shielding even during
the reionization \citep{SU04}, and the star formation can be induced
by SN-driven shock \citep{Mori04}.

As for the reionization, it has been found that the UV emission from
Pop III stars can make a noticeable contribution to the early
reionization. If the escape fraction of ionizing photons is large
enough, then the UV emission from Pop III stars can keep ionizing the
IGM at $7 \leq z \leq 12$.  But, if it is smaller, it can lead to the
re-neutralization of IGM at some redshift, which may result in the
double reionization \citep{Cen03b,Wyi04}.  Regarding the enrichment, a
high Pop III fraction in high-$z$ GRBs may lead to the overproduction
of IGM metals if all synthesized metals are spread out into the
intergalactic space. But, if they are confined around Pop III objects,
it may be difficult to account for IGM metals solely by Pop III stars.

  Finally, it is noted that there still remain a few less robust
factors in this type of analysis. The most unknown factor is the
validity of the spectral peak energy-to-luminosity relation for GRBs
out to $z \sim $ 12.  Although Yonetoku et al. (2004) derived a firm
relation out to about $z \sim $ 5 based on the observed redshifts,
this relation to higher redshifts is not confirmed.  The second key is
the correction for collimation of GRBs.  GRBs are really collimated,
but we do not know a degree of collimation in luminosities for GRBs at
$z>5$. These should be confirmed by detecting GRBs out to $z \sim$ 10
in the mission like the {\it SWIFT} GRB satellite.  Also, we should
check the consistency of the obtained Pop III SFR with the infrared
background.

\acknowledgments
 
  We thank the anonymous referee for valuable comments.
This work is supported in part by a grant-in-aid for scientific
research from the Ministry of Education, Science, Culture, Sports and
Technology in Japan for 14204024 (T.M.), 16002003(M.U.) and 05008
(R.Y.).\\

\clearpage

\begin{figure}
\epsscale{.70}
\plotone{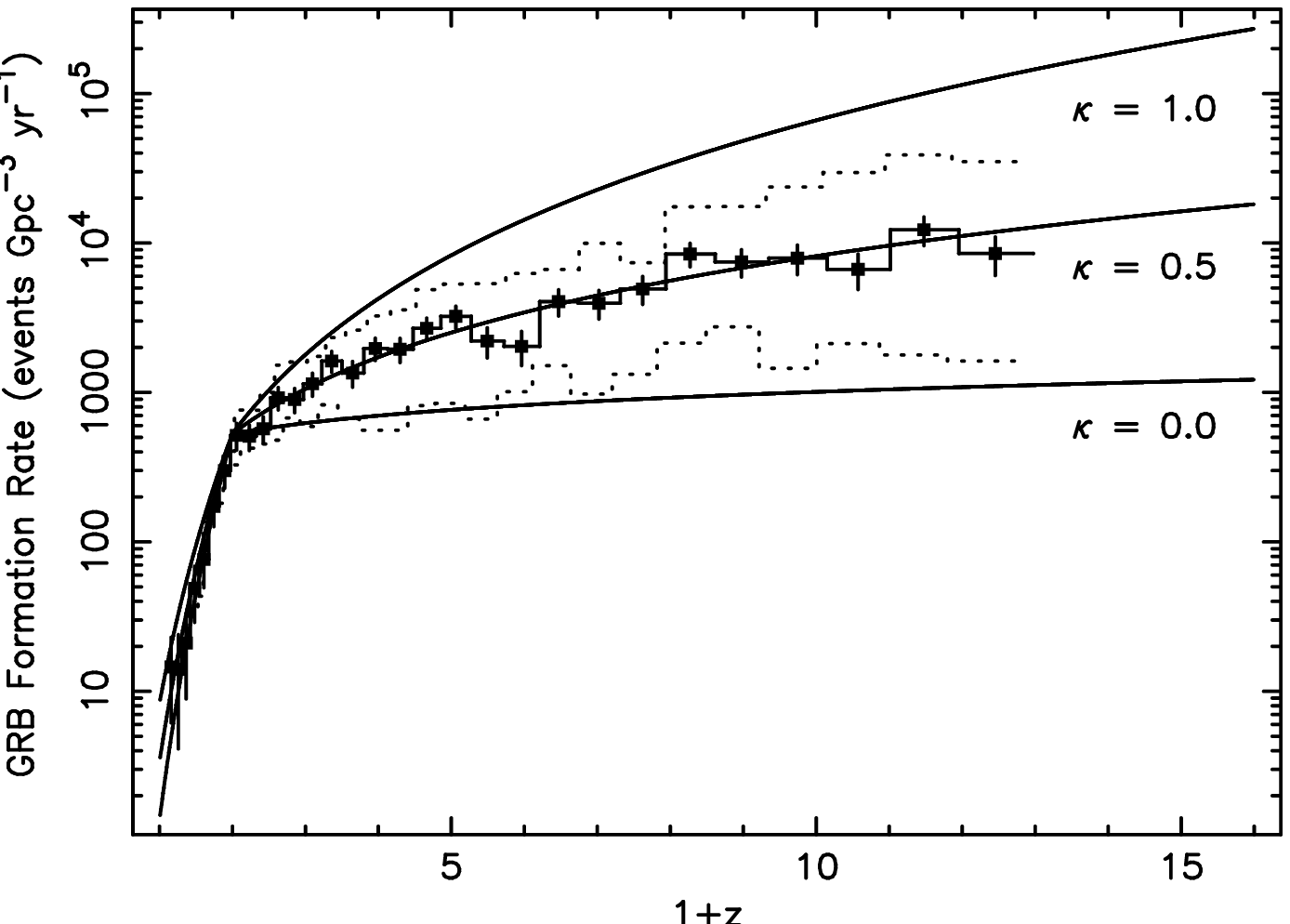}
\caption{ Absolute GRB event rate, $\dot{\rho}_{\rm GRB}$, per unit
cosmological comoving volume, after the jet-collimation evolution
correction.  The emission of GRBs is collimated with typical angle of
0.1 radian \citep{kul99,fra01} and probably shows a cosmological
evolution. We estimate the correction of jet-collimation with a
parameter $\kappa$ as (1+z)${^{2.6\kappa}}$, where 2.6 represents the
observed luminosity evolution.  The squares with error are the
observed GRB event rate after the jet-collimation evolution correction
with $\kappa$=0.5. The raw-data is used in the paper by Yonetoku et
al. \citep{yon04}.  The dotted lines indicate the systematic error of
the formation estimate. The three solid lines show the differences of
the cosmological jet-collimation angle evolution correction with
$\kappa$ = 0, 0.5 and 1.0. The case of $\kappa = 0$ corresponds to no
jet-collimation evolution included. The rough value of
$\dot{\rho}_{\rm GRB}$ is 10${^{4}}$ $Gpc^{-3}$ yr$^{-1}$ at $z \sim
12$. }
\end{figure}

\clearpage

\begin{figure}
\epsscale{.70}
\plotone{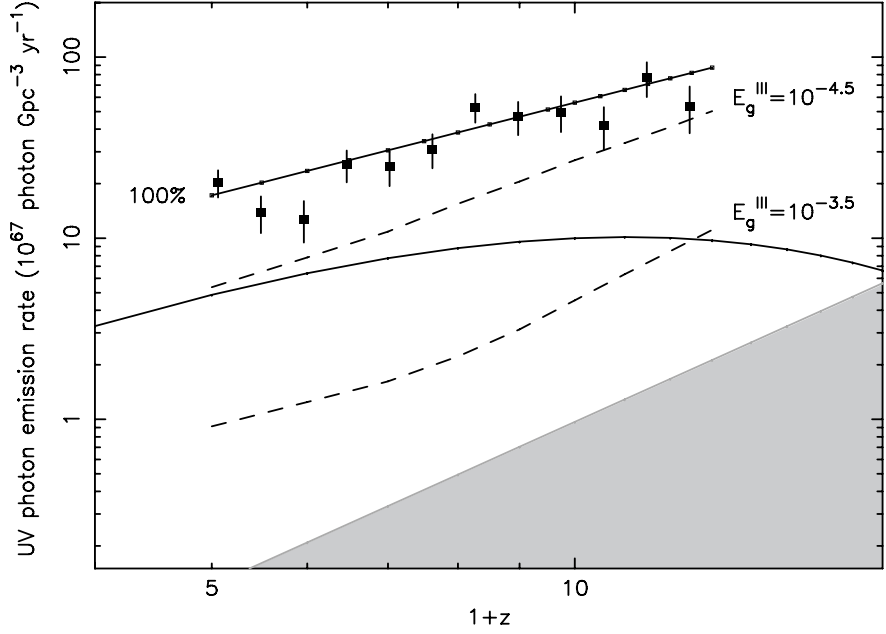}
\caption{ The UV photon emission rate by Pop III stars in a comoving
volume inferred from the Pop III GRB event rate.
A straight line is the emission rate assuming that all GRBs originate
from Pop III stars, where the raw data are plotted with error bars.  
Two dashed lines show the emission rate using the fractions of Pop III 
stars based on the models of \citet{scan03} with $E_g^{\rm III}=10^{-3.5}$ and
$E_g^{\rm III}=10^{-4.5}$, where $E_g^{\rm III}$ is the total energy
input into outfows from Pop III ejecta per unit gas mass. 
The boundary of the gray region shows the minimum level required to 
keep ionizing IGM, assuming the clumping factor of $C=1$.
A solid curve is the maximum level requisite for the IGM ionization,
assuming $C = \langle n_{b}^2 \rangle / \langle n_{b} \rangle^2 $
based on the simulation by \citet{GO97}.
}
\end{figure}

\end{document}